\documentclass[aps,prl,superscriptaddress,twocolumn,preprintnumbers]{revtex4}

\usepackage{graphicx}

\def\be{\begin{equation}}
\def\ee{\end{equation}}
\def\bea{\begin{eqnarray}}
\def\eea{\end{eqnarray}}
\def\met{\slash{\!\!\!{\rm E}}_T}

\begin{document}

\title{Top Quark Forward-Backward Asymmetry and Same-Sign Top Quark Pairs}

\author{Edmond L. Berger}
\email{berger@anl.gov}
\affiliation{High Energy Physics Division, Argonne National Laboratory, Argonne, IL 60439, U.S.A}

\author{Qing-Hong Cao}
\email{caoq@hep.anl.gov}
\affiliation{High Energy Physics Division, Argonne National Laboratory, Argonne, IL 60439, U.S.A}
\affiliation{Enrico Fermi Institute, University of Chicago, Chicago, Illinois
60637, U.S.A.}

\author{Chuan-Ren~Chen}
\email{chuan-ren.chen@ipmu.jp}
\affiliation{Institute for Physics and Mathematics of the Universe, University of Tokyo, Chiba 277-8568, Japan}

\author{Chong Sheng Li}
\email{csli@pku.edu.cn}
\affiliation{Department of Physics and State Key Laboratory of Nuclear Physics
and Technology, Peking University, Beijing 100871, China}

\author{Hao Zhang}
\email{haozhang.pku@pku.edu.cn}
\affiliation{Department of Physics and State Key Laboratory of Nuclear Physics
and Technology, Peking University, Beijing 100871, China}

\begin{abstract}
The top quark forward-backward asymmetry measured at the Tevatron collider shows
a large deviation from standard model expectations.  Among possible interpretations, 
a non-universal $Z^\prime$ model is of particular interest as it naturally predicts a top 
quark in the forward region of large rapidity.   To reproduce the size of the asymmetry, 
the couplings of the $Z^\prime$ to standard model quarks must be large, inevitably 
leading to copious production of same-sign top quark pairs at the energies of the 
Large Hadron Collider (LHC).  We explore the discovery potential for $tt$ and $ttj$ 
production in early LHC experiments at 7-8 TeV and conclude that if {\it no} $tt$ signal is 
observed with 1~fb$^{-1}$ of integrated luminosity, then a non-universal $Z^\prime$ 
alone cannot explain the Tevatron forward-backward asymmetry. 
\end{abstract}

\preprint{ANL-HEP-PR-11-09, EFI-11-4, IPMU11-0008} 

\maketitle
In high energy collisions at the Fermilab Tevatron proton-antiproton collider, 
top quarks are observed to be produced preferentially in the forward hemisphere, 
where forward is defined by the direction of the incident proton beam.
The top quark forward-backward asymmetry $A_{FB}$ 
shows a deviation of three standard deviations ($3~\sigma$) or more from standard 
model (SM) expectations 
in the region of large $t\bar{t}$ invariant mass~\cite{Aaltonen:2011kc}.   
A SM asymmetry in rapidity is predicted from higher order QCD contributions~\cite{Kuhn:1998jr}, 
but it appears to be too small to fit the data. 
Furthermore, a reduction of the asymmetry in $p\bar{p}\to t\bar{t}j$ at next-to-leading order 
is found in Ref.~\cite{Dittmaier:2007wz}.
Several  models 
of new physics (NP) have been invoked to explain the size of the 
asymmetry~\cite{AfbNP,Jung:2009jz,Cao:2009uz,Cao:2010zb,Xiao:2010hm,Choudhury:2010cd}.  A model based on the exchange of a  non-universal massive neutral vector boson 
$Z^\prime$ is intriguing because it naturally 
produces top quarks in the forward region of rapidity via the process $u\bar{u}\to t\bar{t}$,  
with a $Z^\prime$ in the $t$-channel~\cite{Jung:2009jz,Cao:2009uz,Cao:2010zb,Xiao:2010hm,Choudhury:2010cd,Cao:2011ew}.  
This approach requires a flavor changing neutral current (FCNC) interaction $u$-$t$-$Z^\prime$, 
\be
\mathcal{L}= g_W \bar{u}\gamma^\mu (f_L P_L + f_R P_R) t Z^\prime_\mu + h.c., 
\ee
where $g_W$ denotes the weak coupling strength. The left-handed coupling $f_L$ is highly 
constrained by $B_d$-$\bar{B}_d$ mixing: $f_L<3.5\times 10^{-4}~ (m_{Z^\prime}/100~{\rm GeV})$~\cite{Cao:2010zb}.  
We choose $f_L=0$ hereafter.  

\begin{figure}
\includegraphics[clip,scale=0.5]{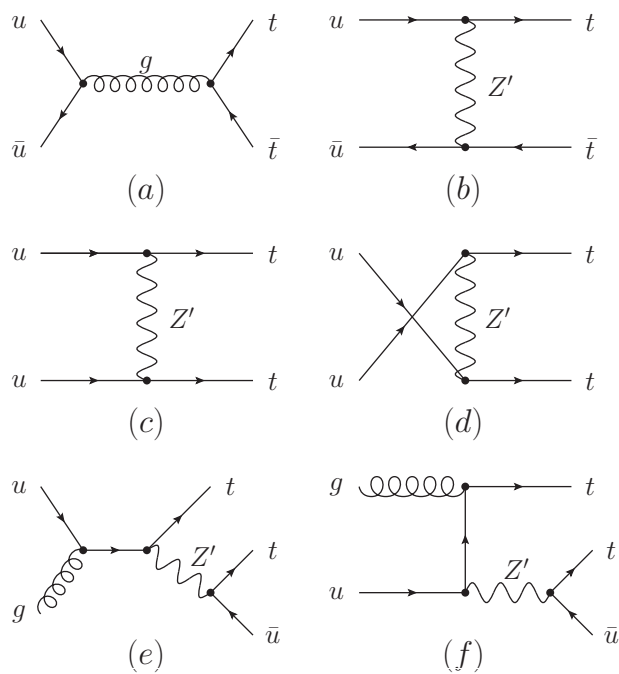}
\caption{Diagrams for (a) $t\bar{t}$ production in the SM, (b) $t\bar{t}$ production
induced by $Z^\prime$ exchange, (c,d) $tt$ pair production, and (e,f) $tt\bar{u}$ 
production.
 \label{feyn}}
\end{figure}

Figure~\ref{feyn}(a) displays the dominant leading-order QCD SM production of a 
$t\bar{t}$ pair at the Tevatron, while Fig.~\ref{feyn}(b) shows $Z^\prime$-induced 
$t\bar{t}$ pair production.  
A NP contribution to  $A_{FB}$ arises from the absolute square of the NP contribution (Fig.~\ref{feyn}(b)) 
and the interference between the NP and the full set of NLO SM QCD amplitudes.  
To produce a large enough asymmetry, 
the coupling $f_R$ must be large if the $Z^\prime$ is heavy ~\cite{Jung:2009jz,Cao:2010zb}.  However, 
it cannot be so large as to result in disagreement with the measured $t \bar{t}$ total cross section
and the $t\bar{t}$ invariant mass distribution.
In this Letter we derive quantitative bounds on $f_R$ and $m_{Z^\prime}$ from Tevatron measurements 
of $A_{FB}$ and the $t \bar{t}$ total cross section, and we use these bounds to predict that  
same-sign $t t$ pair production at the Large Hadron Collider (LHC) should be observed if the 
$Z^\prime$ explanation is correct. 

As illustrated in Fig.~\ref{feyn}(c) and (d), a massive $Z^\prime$ exchange 
inevitably leads to same-sign $tt$ pair production at 
the LHC~\cite{Jung:2009jz, Cao:2010zb, tt}.  The scattering process involves two valence 
$u$-quarks in the initial state and is correspondingly enhanced by the large valence quark 
parton luminosity.   
We focus on the collider phenomenology of  $tt$ pair production in early LHC experiments 
with 7~TeV center-of-mass (c.m.) energy and $1 {~\rm fb}^{-1}$ integrated luminosity.  
In addition to predictions for the rate of same-sign $tt$ pairs,  
we show that the expected right-handed top quark polarization could be measured.  
We further consider same-sign $tt$ pair production in association with a jet, 
as depicted in Fig.~\ref{feyn}(e) and (f), from 
which one can obtain the invariant mass of the $Z^\prime$ from the reconstructed top 
quarks and the additional jet.   Note that there is no resonance in the $tt$ invariant 
mass spectrum since both top quarks are produced in the $t$-channel.   

\begin{figure}
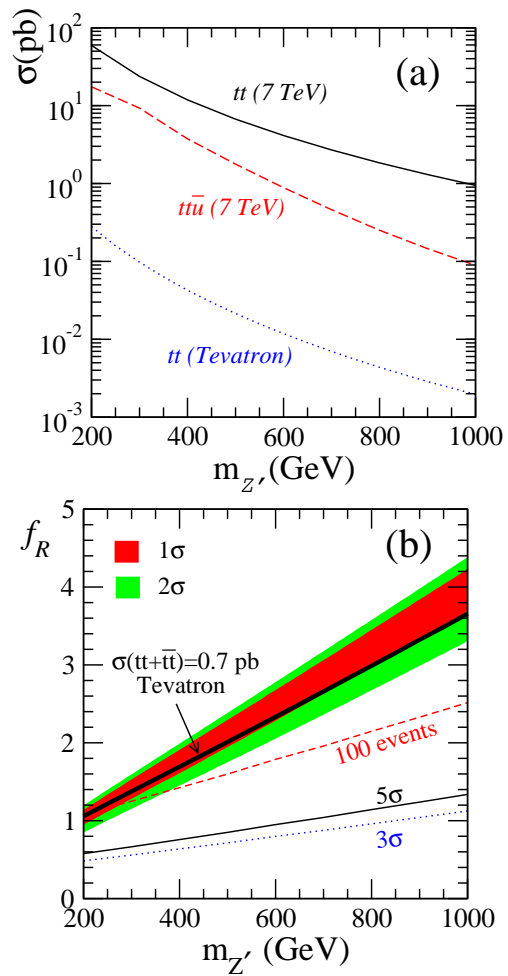

\includegraphics[clip,scale=0.6]{figures/lhc_xsec_new.eps}
\includegraphics[clip,scale=0.6]{figures/lhc_tt_new.eps}
\caption{
a) Inclusive production cross sections for $tt$ and $ttj$ induced by $Z^\prime$ 
exchange, with $f_R=1$, at the LHC (7~TeV) and Tevatron.
(b) The shaded bands in the plane of $m_{Z^\prime}$ and $f_R$ are determined from our 
fit to $A_{\rm FB}$ and $\sigma(t\bar{t})$; the inner (outer) band corresponds to $1\sigma$
($2\sigma$) C.L.  Lines are drawn for $5\sigma$ and $3\sigma$ discovery of $tt$ at the 
7~TeV with an integrated luminosity of $1~{\rm fb}^{-1}$, after all cuts are imposed, as 
specified in the text.  
A dashed line shows the expectation for 100 signal events.   The Tevatron limit on 
$f_R$ from direct search for same-sign top quark pairs is presented.
\label{xsec}}
\end{figure}

In Fig.~\ref{xsec}(a) we display our inclusive cross sections for $tt$ (solid) and 
$tt\bar{u}$ (dashed) as a function of the $Z^\prime$ mass ($m_{Z^\prime}$) at the LHC for $f_R=1$. 
The signal events are generated with MadGraph/MadEvent~\cite{Alwall:2007st}, and
the CTEQ6L parton distribution functions~\cite{Pumplin:2002vw} are used in the calculation.  
We choose the renormalization and factorization scales  to be the top quark mass ($m_t$). 
The $tt\bar{u}$ rate is smaller because it relies on the gluon-quark 
luminosity, smaller than the large valence $uu$ luminosity.   The much smaller rates for 
$\bar{t}\bar{t}$ and $\bar{t}\bar{t}u$ are not shown;  they are suppressed by 
the $\bar{u}\bar{u}$ parton luminosity in a proton-proton collision.  

\begin{table*}
\caption{Signal and background cross sections (fb) for $tt$ pair production 
at the LHC (7~TeV) before and after cuts, with $f_R=1$, 
for nine values of $m_{Z^\prime}$~(GeV) after the restriction to $2\mu^+$'s and with tagging efficiencies included. 
The cut acceptances $\epsilon_{\rm cut}$ are also listed. 
\label{tab:xsec}}
\begin{tabular}{cccc|cccc|cccc|ccccc}
\hline 
$m_{Z^\prime}$ & No cut  &  With cut &  $\epsilon_{\rm cut}$ &
$m_{Z^\prime}$ & No cut  &  With cut &  $\epsilon_{\rm cut}$ &
$m_{Z^\prime}$ & No cut  &  With cut &  $\epsilon_{\rm cut}$ &
Background & No cut &  With cut         &  $\epsilon_{\rm cut}$ &  \tabularnewline
\hline
200  & 730.6  &  72.0 &   9.9\% &500   &  82.8  & 15.3 &  18.5\% & 800   &  22.7  & 4.7   & 20.9\% & $t\bar{t}$    & 1205.2  & 0.4   & 0.03\% \tabularnewline
300  & 292.5  &  41.0 & 14.0\% &600   &  51.0  &   9.8 &  19.3\% & 900   &  16.1  & 3.4   & 21.2\% & $WWjj$        &   115.8  & 0.2   & 0.16\%   \tabularnewline
400  & 146.4  &  24.3 & 16.6\% &700   &  33.3  &   6.8 &  20.4\% &1000  &  11.7  & 2.5   & 21.2\% & $WWW/Z$   &       0.4  & 0.01 & 2.5\%   \tabularnewline
\hline
\end{tabular}
\end{table*}

In order to trigger on same-sign $tt$ events, we demand that 
both top quarks decay leptonically and we further concentrate on the $\mu^+$ as its charge
can be better determined~\cite{Aad:2009wy}.  
Needless to say, including the electrons 
would improve the discovery potential. The sample of events of interest to us is defined by 
$\mu^+\mu^+~ b~b~\met$, 
where the missing transverse momentum $\met$ originates from two unobserved neutrinos.  
Our procedure for simulating the signal and background processes at the parton level, retaining all spin correlations, is similar to that described in Refs.~\cite{Berger:2010fy,Zhang:2010kr}, to which we refer readers for details.  The dominant SM backgrounds are:
\bea
pp & \to & W^{+}(\to \ell^+\nu)W^{+}(\to \ell^+\nu)jj, \\
pp & \to & t\bar{t} \to bW^{+}(\to \ell^+\nu)\bar{b}(\to \ell^+)W^{-}(\to jj), 
\eea
computed with ALPGEN~\cite{Mangano:2002ea}.  
Other SM backgrounds, e.g. triple gauge boson production 
($WWW$, $ZWW$, and $WZg(\to b\bar{b})$), 
occur at a negligible rate after kinematic cuts.  
Since muon charge identification is not perfect,  we remark that  
$t\bar{t}$ pair production could also be a background when $\mu^-$ leptons from 
the antitop quark decay are misidentified as $\mu^+$ leptons. However,  
this background is negligible~\cite{Zhang:2010kr}.

At the analysis level, all signal and background events are required to pass the
following acceptance cuts:
\bea
&& n_j = 2,~n_{\mu^+} = 2,~p_T^j\geq 50\,{\rm GeV}, ~\left|\eta_{j}\right|\leq 2.5,~\nonumber \\
&& p_{T}^{\ell}\geq 50\,{\rm GeV},\quad \left|\eta_{\ell}\right|\leq2.0, \quad \met>20~{\rm GeV},\nonumber \\
&&\Delta R_{jj,j\ell,\ell\ell} > 0.4,
\label{eq:cut}
\eea
where the separation $\Delta R$ in the azimuthal angle ($\phi$)-pseudorapidity ($\eta$) plane between the objects $k$ and $l$ is
$\Delta R_{kl}\equiv\sqrt{\left(\eta_{k}-\eta_{l}\right)^{2}+\left(\phi_{k}-\phi_{l}\right)^{2}}$.
The two jets are further required to be $b$-tagged.
We also model detector resolution effects as described in Ref.~\cite{Zhang:2010kr}.

Table~\ref{tab:xsec} shows the signal and background cross sections (in fb units) for $tt$ pair production
before and after cuts, with $f_R=1$, for nine values of $m_{Z^\prime}$. The rates for other values of $f_R$ can be obtained from: 
\be
\sigma(tt) = \sigma_{f_R=1}(tt) \times f_R^4.
\ee
The SM backgrounds are suppressed efficiently such that less than 1 background event survives after cuts with an integrated luminosity of $1~{\rm fb}^{-1}$. 
Based on Poisson statistics, one needs 8 signal events 
in order to claim a $5\sigma$ discovery significance on top of 1 background event. 
The discovery potential is plotted in Fig.~\ref{xsec}(b) with  black-solid  ($5\sigma$) and blue-dotted ($3\sigma$) curves.  

The forward-backward rapidity asymmetry $A_{\rm FB}$ is defined as 
\begin{eqnarray}
A_{\rm FB}^{\rm tot} & = &  \frac{\sigma_F-\sigma_B}{\sigma_F + \sigma_B} =
 \frac{\sigma_{F}^{\rm SM}-\sigma_{\rm B}^{\rm SM}+\sigma_{\rm F}^{\rm NP}-\sigma_{\rm B}^{\rm NP}}
      {\sigma_{F}^{\rm SM}+\sigma_{\rm B}^{\rm SM}+\sigma_{\rm F}^{\rm NP}+\sigma_{\rm B}^{\rm NP}}
       \label{eq:AFB1} \nonumber \\
 & = &
 \frac{\sigma_{\rm F}^{\rm NP}-\sigma_{\rm B}^{\rm NP}}{\sigma_{\rm F}^{\rm NP}+\sigma_{\rm B}^{\rm NP}}
 \times
 \left(1+
  \frac{\sigma_{\rm F}^{\rm SM}-\sigma_{\rm B}^{\rm SM}}{\sigma_{\rm F}^{\rm NP}-\sigma_{\rm B}^{\rm NP}}
 \right)
 \times
 \frac{\sigma_{\rm tot}^{\rm NP}}{\sigma_{\rm tot}^{\rm SM}+\sigma_{\rm tot}^{\rm NP}}\nonumber \\
 & = & 
 A_{\rm FB}^{\rm NP}\times R+A_{\rm FB}^{\rm SM}\left(1-R\right)
 \label{eq:AFB}
\end{eqnarray}
where
\bea
A_{\rm FB}^{\rm NP} &\equiv&
(\sigma_{\rm F}^{\rm NP}-\sigma_{\rm B}^{\rm NP})/(\sigma_{\rm F}^{\rm NP}+\sigma_{\rm B}^{\rm NP}),
\nonumber \\
A_{\rm FB}^{\rm SM} &\equiv&
(\sigma_{\rm F}^{\rm SM}-\sigma_{\rm B}^{\rm SM})/(\sigma_{\rm F}^{\rm SM}+\sigma_{\rm B}^{\rm SM})
\nonumber \\
R&\equiv&(\sigma_{\rm tot}^{\rm NP})/(\sigma_{\rm tot}^{\rm SM}+\sigma_{\rm tot}^{\rm NP}) 
\label{eq:def}
\eea
are the asymmetries induced by NP and in the SM,
and $R$ is the fraction of the NP contribution
to the total cross section. 
Here, $\sigma_{F(B)}$ denotes the $t\bar{t}$ cross section in the forward (F) and backward (B) rapidity region.
The standard model QCD and new physics contributions to the cross sections 
are denoted by superscripts SM and NP.  

The shaded regions in the $f_R$ plane in Fig.~\ref{xsec}(b)  are derived from requiring consistency with both $A_{FB}$~\cite{Aaltonen:2011kc} and the $t\bar{t}$ production cross section $\sigma(t\bar{t})$~\cite{CDF:9913}:
\bea
A_{\rm FB} & = & 0.475 \pm 0.114~~{\rm for}~~m_{t\bar{t}}\geq 450~{\rm GeV} \nonumber \\
\sigma(t\bar{t}) & = &7.50 \pm 0.48~{\rm pb}.
\eea
The inner (red) and outer (green) regions correspond to $1\sigma$ and $2\sigma$ C.L., respectively. 
The SM predictions of  $A_{\rm FB} (m_{t\bar{t}}\geq 450~{\rm GeV})$ and  $\sigma(t\bar{t})$ calculated with $m_t = 172.5~{\rm GeV}$ are $0.088$~\cite{Aaltonen:2011kc} and $6.9~{\rm pb}$~\cite{Cao:2010zb}, respectively.
The lower bound of each band is derived from the $A_{\rm FB}$ measurement 
while the upper bound is from the $\sigma(t\bar{t})$ data.  
In addition we verify that our computed distribution in $m_{t\bar{t}}$ is consistent with 
recent CDF data~\cite{Aaltonen:2009iz} at the level of $\lesssim 2\sigma$ deviations.

The search for same sign top quark pairs at the Tevatron, 
$\sigma(tt+\bar{t}\bar{t}) \lesssim 0.7~{\rm pb}$~\cite{Aaltonen:2008hx},  
imposes a constraint on $f_R$ and $m_{Z'}$ shown by the black band in Fig.~\ref{xsec}(b). 
Parts of the otherwise allowed $1\sigma$ and $2\sigma$ bands are excluded by these data.

The values of $f_R$ indicated by the shaded bands in Fig.~\ref{xsec}(b) show that $f_R \gtrsim 1$ for 
all $m_{Z^\prime}$. They are everywhere above the 
values needed for $5$ standard deviation observation of same sign $t t$ pair production at the LHC.  
We conclude that if {\it no} $tt$ signal is 
observed with 1~fb$^{-1}$ of integrated luminosity at the LHC, then a non-universal $Z^\prime$ 
alone cannot explain the Tevatron forward-backward asymmetry. 

\begin{figure}
\includegraphics[clip,scale=0.57]{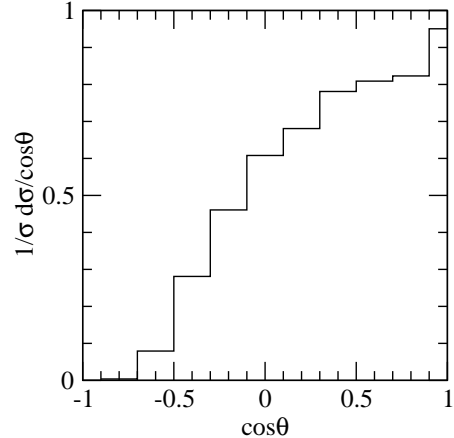}
\caption{Normalized distribution of the angle of the charged lepton relative to the top quark in the c.m. 
frame in the $tt$ pair production after cuts and efficiencies are included for 
$m_{Z^\prime}=800~\rm{GeV}$ and $f_R=1$.
\label{distro}}
\end{figure}

If an excess is observed in the $\mu^+\mu^+bb$ plus $\met$ sample,  one must demonstrate consistency with a $uu\to tt$ origin.  
Top quark polarization is a good probe of the 
FCNC $Z^\prime$ because the right-handed $u$-$t$-$Z^\prime$ coupling forces the top quarks 
to be mainly right-handed polarized.   Reconstructing the two top quarks and measuring their polarizations would permit validation of the FCNC $Z^\prime$ model. 
Among the top quark decay products the charged lepton is maximally correlated with the top quark spin. 
In our signal process the charged lepton from  top quark decay exhibits a $1+\cos\theta$ distribution, 
where $\theta$ is the helicity angle between the charged lepton momentum in the top quark rest frame and top quark momentum in the
c.m. frame of the production process. 
Following Ref.~\cite{Berger:2010fy}, we use the MT2 method~\cite{Lester:1999tx}
to select the correct $\mu$-$b$ combinations and to verify whether the final state is 
consistent with $t \to Wb$ parentage.   
Then we make use of the on-shell
conditions of the two $W$ bosons and the two top quarks to solve for 
the neutrino momenta~\cite{Sonnenschein:2006ud,Bai:2008sk}.  
Once the neutrino momenta are known, the kinematics of the entire final state are fixed and 
the angular distribution may be constructed.

The reconstructed $\cos\theta$ distribution after cuts is plotted in Fig.~\ref{distro}, 
and it clearly shows the expected $1+\cos\theta$ form.
The discovery potential of the $tt\bar{u}$ signature is also promising.  If a peak can be 
found in the invariant mass spectrum of a $t$ and a  light jet 
(from the $\bar{u}$ in Fig.~\ref{feyn}(e) and (f)), 
one could confirm the presence of the FCNC $Z^\prime$.

\begin{acknowledgments}
The work by E.L.B. and Q.H.C. is supported in part by the U.S. DOE
under Contract No.~DE-AC02-06CH11357. Q.H.C. is also
supported in part by the Argonne National Laboratory and University
of Chicago Joint Theory Institute Grant 03921-07-137. C.R.C. is supported by World Premier
 International Center Initiative (WPI Program), MEXT, Japan.
C.S.L. and H.Z. are supported in part by the National Natural Science Foundation of China, under Grants No.11021092 and No.10975004.
Q.H.C. thanks Shanghai Jiaotong University for hospitality where part of this work was done. 
E.L.B. and C.R.C. also thank NCTS in Taiwan for hospitality where part of his  work was done.
\end{acknowledgments}

\end{document}